\documentclass[conference]{IEEEtran}

\usepackage{cite}
\usepackage{amsmath,amssymb,amsfonts}
\usepackage{algorithmic}
\usepackage{graphicx}
\usepackage{textcomp}
\usepackage{xcolor}
\def\BibTeX{{\rm B\kern-.05em{\sc i\kern-.025em b}\kern-.08em
    T\kern-.1667em\lower.7ex\hbox{E}\kern-.125emX}}

\begin{document}
\bstctlcite{IEEEexample:BSTcontrol}
\title{Application of Stochastic Optimization Techniques to the Unit Commitment Problem --- A Review}

\author{\IEEEauthorblockN{Vincent Meilinger}
\IEEEauthorblockA{\textit{Technische Universität Berlin} \\
Berlin, Germany \\
v.meilinger@campus.tu-berlin.de}}

\maketitle

\begin{abstract}
Due to the established energy production methods contribution to the climate crisis, renewable energy is to replace a substantial part of coal or nuclear plants to prevent greenhouse gases or toxic waste entering the atmosphere. This relatively quick shift in energy production, primarily pushed by increasing political and economical pressure, requires enormous effort on the part of the energy providers to balance out production fluctuations. Consequently, a lot of research is conducted in the key area of stochastic unit commitment (UC) on electrical grids and microgrids. The term unit commitment includes a large variety of optimization techniques, and in this paper we will review recent developments in this area. We start by giving an overview over different problem definitions and stochastic optimization procedures, to then assess recent contributions to this topic. We therefore compare the proposals and case studies of several papers.
\end{abstract}


\section{Introduction}
Since the advent of renewable energy sources the need for scheduling power production and energy distribution has become more important than ever. The energy market is shifting from inflexible power production with coal or nuclear power plants to the highly fluctuant wind or solar energy plants that require more flexible control mechanisms \cite{zheng2015}. Especially grid balancing, which includes controlling available generation units and prediction of energy production and load, is an increasingly difficult task, due to the stochastic nature of renewable resources \cite{nguyen2016}. The process used by power generation companies to create predictions about the energy production needs days or weeks ahead of time is called unit commitment (UC) \cite{zheng2015}. The actual real time allocation of units minutes or hours before is referred to as economic dispatch (ED). The focus of this review lies on the prior research topic.

UC decisions define, for each time step over a scheduling horizon, which generation unit should be used, while accounting for market clearing, reliability assessment, intra-day operations, expected load and operational policies \cite{zheng2015}, \cite{yurdakul2020}.
Energy grids mainly comprise multiple distributed generation resources, electric storage resources and loads. \cite{yurdakul2020_quantification}. Resource generation units can generally divided into several groups: thermal generation resources (TGR) like coal or nuclear, renewable resources (RR) like wind and solar, and resource storage, like batteries or gas storage. 
There are many groups of optimization algorithms that try to solve the complex production and demand balancing. \cite{zheng2015} 
Since many parameters of energy production are uncertain, especially with wind and solar, stochastic optimization methods are a popular choice to optimize for cost, reliability, equipment failure and more \cite{haberg2019}.

UC therefore represents a optimization problem under constraints and uncertainty. UC is proven to be an NP-hard problem \cite{zheng2015}. In this review we compare three approaches to solve the UC problem: stochastic programming (SP), stochastic dynamic programming (SDP) and distributionally robust optimization (DRO). 
The paper is structured as follows.\\
In Section \ref{SO_problems} we define the basic objective of SO problems in UC. We summarize various approaches to solve UC.
Thereafter, in Sections \ref{DRO_section}, \ref{SP_section} and \ref{SDP_section}, we review different DRO, SP and SDP models, respectively. To evaluate performance, the case studies are analyzed.
Section \ref{comparison_section} the advantages and disadvantages of the optimization techniques are evaluated.
A short discussion regarding future development is held in Section \ref{outlook_section}, with a focus on incorporating green house gas emissions in UC.

In the next section we give an overview over the different UC modeling and solving approaches.

\section{Stochastic optimization problems} \label{SO_problems}
The term 'stochastic optimization' includes a variety of optimization techniques. When applied to UC for power grids, the problem can be defined as an optimization of an objective function under a set of constraints \cite{nguyen2016}. Underlying the function parameters are probability distributions, which in case of stochastic programming are assumed to be known prior to realization \cite{yurdakul2020}.

To formalize the problem, we need to give an overview of suitable methods to model the uncertainty of the problem first.
UC models are often defined as two-stage models. The first stage contains commitment decisions $\textbf{u}$, while dispatch decisions are incorporated in the second stage after the uncertain parameters are realized \cite{haberg2019}. In two stage models the objective function is often quadratic or piecewise linear \cite{haberg2019}. Uncertainty is realized only once \cite{zheng2015}.

Multistage models also exist, where commitment decisions are modeled as nodes of a scenario tree \cite{haberg2019}. Uncertainty can therefore be realized over time, decisions can hence be adjusted dynamically \cite{zheng2015}. Therefore, these models define the uncertain future as a scenario tree that includes many possible simulated outcomes \cite{zheng2015}. This method is common for stochastic programming solutions. The scenario tree can be searched using Monte Carlo simulations \cite{nguyen2016}. A disadvantage of this technique is the exponentially increasing computational complexity when growing the number of modeled scenarios \cite{nguyen2016}. In \cite{haberg2019}, algorithms used to cope with these enormous scenario trees are listed, among them for example Benders decomposition, Lagrangian relaxation, dual decomposition or Progressive hedging as used in \cite{liu2021}. The latter approach will be explored in more detail later on.
It is also possible to use uncertainty sets or probabilistic constraints. Uncertainty sets are often used to model uncertainty for robust optimization models \cite{zheng2015}. Confidence intervals are the most basic uncertainty sets, they are defined by the sample mean and quantiles of an according distribution \cite{chen2018}. 
A general definition of some SP solving approaches can be found in \cite{zheng2015}.
The authors show that stochastic programming models often take the form 
\begin{equation} \label{eq:SO_def}
    \underset{\mathbf{u} \in \mathbb{U}}{\text{min}} \; \mathbf{c}^\intercal\mathbf{u} \; + \; E_\xi [F(\mathbf{u}, \xi)] 
\end{equation}
where $\mathbf{u}$ are the commitment decisions of slow reacting power units like coal and nuclear and $\mathbb{U}$ is the set of feasible commitment decisions, startup and shutdown costs are represented by the vector $\mathbf{c}$ and $\xi$ is a vector of uncertain parameters drawn from a known probability distribution \cite{zheng2015}. The first term therefore models day-ahead planning, which includes rather unresponsive commitment decisions e.g. of TGR generators like coal and nuclear, while the second term models real time decisions \cite{zheng2015}. The latter ED phase is needed to adjust for fluctuations of renewable energy production or outages.
While stochastic programming models have the advantage of cost saving and reliability improvements when compared to simply using reserve constraints \cite{zheng2015}, a major drawback is the assumption that the distribution of the data points used for the model is known a priori \cite{yurdakul2020}. In realistic scenarios, this is often not the case.  

Another wide spread technology is robust optimization (RO) \cite{chen2018}. Robust optimization models uncertainty through uncertainty sets. This method bases its decisions on the worst case outcome, its predictions are therefore rather conservative \cite{chen2018} \cite{yurdakul2020}. 
The general robust optimization problem is defined as 
\begin{equation} \label{eq:RO_def}
    \underset{\mathbf{u} \in \mathbb{U}}{\text{min}} \left\{ \mathbf{c}^\intercal\mathbf{u} \; + \; \underset{\mathbf{v} \in \mathbb{V}}{\text{max}} [F(\mathbf{u}, \mathbf{v})]\right\}
\end{equation}
where $\mathbf{u}$ and $\mathbb{U}$ are defined analogous to (\ref{eq:SO_def}), $\mathbf{v}$ again represents the uncertainty parameter with deterministic uncertainty set $\mathbb{V}$ \cite{zheng2015}. The function $F(\mathbf{u}, \mathbf{v})$ models the real-time dispatch cost for the decisions $\mathbf{u}$ and an uncertain variable $\mathbf{v}$ \cite{zheng2015}. The difference between the vectors $\mathbf{v}$ of SO and RO is that while for SO the vector represents a vector of uncertain parameters, for RO it is a variable vector. For the latter, $\mathbf{v}$ needs to be adjusted to minimize the worst scenarios total cost \cite{zheng2015}. 

In both SP and RO models second stage problem, function $F$ represents a minimization task. The problem definition is very similar in both methods, as shown in \cite{zheng2015}, where the general SP function is defined as 
\begin{align} \label{eq:ss_SO}
    &F(\mathbf{u}, s) = \underset{\textbf{p}_s, \textbf{f}_s}{\min} \; f(\textbf{p}_s)\\
    &\text{s.t. } A_\textbf{s} \textbf{u} + B_\textbf{s} \textbf{p}_s + H_\textbf{s} \textbf{f}_s \geq \textbf{d}_\textbf{s}
\end{align}
for each realization $s$ of random vector $\xi$, and the RO function as
\begin{align} \label{eq:ss_RO}
    F(\mathbf{u}, \mathbf{v}) =& \underset{\textbf{p, f}}{\min} \; \textbf{q}^\intercal \textbf{p} \\
            & \text{s.t. } A_\textbf{v} \textbf{u} + B_\textbf{v} \textbf{p} + H_\textbf{v} \textbf{f} \geq \textbf{d}_\textbf{v}
\end{align}

In (\ref{eq:ss_SO}) Matrices $A_s, B_s, H_s$ usually model contingencies, while $\textbf{d}_s$ models uncertain demand and renewable energy outputs \cite{zheng2015}. Vectors $\textbf{p}_s, \textbf{f}_s$ model dispatches and reserves of multiple periods and other second stage decisions, respectively \cite{zheng2015}. Function $f$ models fuel cost and is usually quadratic and non-convex. The variables and parameters are similarly defined in (\ref{eq:ss_RO}), with the only difference that the matrices and vectors are functions of the uncertain parameters, and the objective function uses piecewise linear approximation with coefficient vector $\textbf{q}$ to approximate the original SO cost function $f$ \cite{zheng2015}.

A newer optimization method, the distributionally robust optimization (DRO) could solve the problems of RO and SO \cite{yurdakul2020}, \cite{chen2018}. DRO takes into account the shortcomings of RO, in that it considers the uncertainty of the parameter probability distribution \cite{yurdakul2020}. The probability distributions here belong to ambiguity sets that are constructed from historical data using methods like Kullback-Leibler divergence, moment information or the Wasserstein distance \cite{yurdakul2020}.

SDP solves the problem by breaking it down into a sequence of steps over time \cite{nguyen2016}. In \cite{zheng2015}, a general finite-horizon, discrete-time SDP framework is defined as follows:
\begin{equation}
    \underset{\pi \in \Pi}{\text{inf}} \; V_\pi(s_0) := \mathbb{E}\left[ \underset{t=0}{\overset{T-1}{\sum}} C_t(s_t, \mu_t(s_t), \xi_t) + C_T(u_T) \right].
\end{equation}
Here, $\xi_t$ are random variables, $C_t(\cdot)$ models system cost at time period $t$, and $V_\pi$ is referred to as the value function. $s_t$ is the state of a system at time period $t$, $u_t$ are the commitment and dispatch decisions according to policy function $\mu(\cdot)$ which maps from system state $s_t$
At each time step, the best route is found based on the possible optimum subsequences in the previous step \cite{nguyen2016}. The main advantage of dynamic programming (DP) is that it can maintain the feasibility of the solution, as it is able to find the optimal subsequence while searching for the optimal sequence \cite{nguyen2016}.

In the following sections we will summarize several papers using the previously listed methods.

\section{Distributionally robust optimization} \label{DRO_section}
In \cite{yurdakul2020}, the authors present a distributionally robust optimization (DRO) technique that minimizes the expected cost from the worst case scenario over an ambiguity set of probability distributions constructed by Kullback-Leibler (KL) divergence. 

Kullback-Leibler divergence is a measure of distance between probability distributions \cite{chen2018}. The nominal distribution $P_0$ can be constructed using parametric methods like maximum likelihood estimators or non-parametric methods. The latter are often used to handle data driven problems, because they do not assume anything about the underlying datapoint distribution \cite{chen2018}.

The DRO approach is applied to microgrids and incorporates the evaluation of microgrid net load uncertainty and electricity market prices \cite{yurdakul2020}.
The proposed approach is modeled as a two-stage problem, where the first stage determines the binary commitment and startup variables of the thermal generation resources (TGR) over a set period, incorporating typical TGR constraints like uptime and downtime \cite{yurdakul2020}. The uncertain net load and the electricity price are not realized yet, since the objective of the first stage is to minimize fixed costs and worst case expected power costs.
The second stage problem minimizes the power generation and purchase costs. It is constrained by the TGR output limits, up/-downtime constraints and power balancing.
The authors decided to model the uncertainty through ambiguity sets, which is a common choice for RO models \cite{zheng2015}. 

The ambiguity set $\mathbf{P}$ in \cite{yurdakul2020} is constructed with KL divergence through k-means clustering and is defined as
\begin{align}
\mathbf{P} := \{\mathcal{P}: & \underset{\omega = 1}{\overset{S}{\sum}} \pi^\omega_\mathcal{P} \log \left( \frac{\pi^\omega_\mathcal{P}}{\pi^\omega_{\mathcal{P}_o}} \right) \leq \rho, \label{eq:KLdiv1}\\
&\underset{\omega = 1}{\overset{S}{\sum}} \pi^\omega_\mathcal{P} = 1, \label{eq:KLdiv2} \\
&\pi^\omega_\mathcal{P} \geq 0 \;\;\; \forall\omega \in \Omega\}. \label{eq:KLdiv3}
\end{align}
Divergence tolerance $\rho$ is used to adjust the ambiguity set size and therefore its conservatism \cite{yurdakul2020}. $\omega$ represents a cluster, cluster centroids represent realizations $\xi^\omega$. Scenario $\omega$ has nominal probability $\pi^\omega_{\mathcal{P}_o}$ for the nominal probability distribution $\mathcal{P}_o$. The number of scenarios is denoted as $\mathcal{S}$.
The resulting KL divergence based microgrid UC (KL-MUC) is reformulated to a convex mixed integer non-linear problem (RKL-MUC) defined as follows:
\begin{align} \label{RKL-MUC}
    \text{RKL-MUC}: &\\
    \underset{x, \mu, \zeta}{\text{minimize}} \;\;\; & \;\; c \cdot x + \mu + \rho\zeta + \zeta  \underset{\omega = 1}{\overset{\mathcal{S}}{\sum}} \pi^\omega_{\mathcal{P}_o} e^{\overline{\mathcal{K}}^\omega (x, \mu, \zeta)-1}\\
    s.t. \;\;\; & \;\; x \in \mathbf{X}, \\
                & \;\; \zeta \geq 0,
\end{align}
where $\mathbf{X}$ represents the feasibility region of x defined by the first stage problem constraints, and $\overline{\mathcal{K}}^\omega (x, \mu, \zeta) = \frac{\mathcal{Q}(x, \xi^\omega)-\mu}{\zeta}$ \cite{yurdakul2020}. $\mathcal{Q}(x, \xi^\omega)$ is a function of the uncertain power generation and purchase cost for variable $x$ and realization $\xi^\omega$ of uncertain values \cite{yurdakul2020}.
Dual variables $\zeta, \mu$ are are assigned to the constraints (\ref{eq:KLdiv1}) and (\ref{eq:KLdiv2}), respectively \cite{yurdakul2020}. Constraint (\ref{eq:KLdiv1}) controls the divergence between $\mathcal{P}_o$ and distribution $\mathcal{P}$ incorporated into ambiguity set $\textbf{P}$ based on KL divergence, (\ref{eq:KLdiv2}) ensures the sum of probabilities assigned by $\mathcal{P}$ is one \cite{yurdakul2020}.
The RKL-MUC problem is solved using an algorithm based on Benders decomposition, where it is decomposed into a lower bounding master problem and an upper bounding subproblem \cite{yurdakul2020}.
The approach enables hedging against adopting misrepresenting probability distributions \cite{yurdakul2020}. When benchmarked against a SUC problem equivalent to the formulation of KL microgrid unit commitment problem that is solved by the L-shaped algorithm, the total cost of RKL-MUC is observed to be less than or equal to the cost under the SUC formulation \cite{yurdakul2020}. Furthermore, while both problem formulations give similar results when divergence parameter $\rho = 0$, the total cost of the RKL-MUC solution gets smaller compared to the SUC solution. The authors state that this result can be traced back to the fact that it is beneficial to take additional probability distributions other than the nominal distribution into account \cite{yurdakul2020}.

Later in \cite{yurdakul2021}, Yurdakul et al. further build on their previous research in distributionally robust unit commitment under net load uncertainty using KL divergence. Improvements in the short term planning are achieved by reformulation of the KL distrubutionally robust unit commitment (KL-DRUC), obtaining a convex mixed-integer nonlinear reformulated KL-DRUC (RKL-DRUC) problem where the objective function is defined in a similar way as the objective function  (\ref{RKL-MUC}) of the previous paper, with the difference of more constraints incorporated and the focus shifted from microgrids to power systems under net load uncertainty in general \cite{yurdakul2021}. 
Using a similar SUC model as the previous paper, the focus here lies on investigating the influence of different k-means clustering distance measures, the size of the dataset used for ambiguity set construction and the influence of different values of divergence parameter $\rho$ on the results \cite{yurdakul2021}. Furthermore, more constraints are incorporated into the second stage, namely the load curtailment cost, ramping constraints \cite{yurdakul2021}. 
The authors this time present the solution in the form of a real world data case study, where net load values of the California System Operator Grid are used. To gain an insight into the impact of $\rho$, the authors optimize the model for different values and different distance functions. A increase in total cost can be observed for larger values of $\rho$ for all tested distance measures. Furthermore, a coupling between optimal cost and number of datapoints can be noted, in that the total cost decreases with an increasing count of time-series data points.
The research provided in this paper could aid grid operators with the UC decision process for grids with an increasing amount of renewable energy generators.

\section{Stochastic Programming} \label{SP_section}
A classical SP approach can be found in \cite{hanhuawei2017}. The authors provide a method to integrate stochastic solar power into power grids. The fist stage solves the problem of UC and economic dispatch based on predicted solar power generation. The second stage handles thermal generation rescheduling when solar power is generated. In order to cope with overestimation of solar production, additional reserve purchasing is included in the model. All in all, the model is constrained by power balance, minimum/maximum generation, operating reserve, ramp rate, up-/downtime,  operating costs and reserve buying/penalty costs. The objective function to minimize is defined as
\begin{align}
    \min \underset{t=1}{\overset{T}{\sum}} (S_t + & \overline{E}_S(O_{s, t}(\Tilde{p}_{pv, s, t}, \Tilde{\mathbf{p}}_{s, t}, \textbf{u}_t)) + \\
    & \overline{E}_S(\omega_t |\Tilde{p}_{pv, s, t} - \Tilde{P}_{pv, s, t}|)),
\end{align}
where $\textbf{u}_t$ are the UC decisions of thermal units, $\Tilde{\textbf{p}}_t, \Tilde{p}_{pv, s, t}$ are the scheduling decisions depending on the actual and stochastic solar power generation. $\overline{P}_{pv, s, t}$ is a stochastically simulated scenario and decision variables with subscript $s$ are the corresponding scheduling decisions in a scenario $s$. $\overline{E}_S$ is an approximation to expected value $E$ with S solar power scenarios. $S_t$ is the startup cost.
The model is tested on a simulation containing several loads and generators, solar and thermal \cite{hanhuawei2017}. When comparing the proposed method to deterministic or robust optimization, a significant improvement can be noticed \cite{hanhuawei2017}. With all three methods, the average cost rises with rising solar uncertainty due to over-scheduling or penalty cost, but the stochastic approach stays relatively constant in comparison to the other models \cite{hanhuawei2017}. Especially the deterministic model suffers from this uncertainty \cite{hanhuawei2017}. The robust model is only slightly worse than the stochastic model, because its decisions are based on the worst case scenario and, as mentioned earlier, therefore rather conservative \cite{hanhuawei2017}.

When using ambiguity sets constructed from observed real world data as in \cite{yurdakul2021}, DRO still has a main advantage over SP methods. SP can potentially suffer from an inaccurate distribution of uncertain parameters \cite{chen2018}. Of course, even the distribution of the observed data in the ambiguity set can differ from the underlying real distribution when measurement errors occur, but in reality the obtained distribution is often close to the ground truth \cite{chen2018}. The disadvantage of distance based DRO is the high complexity of the problem, as the objective function of these problems are often non-linear \cite{chen2018}.
To counteract this problem, in \cite{chen2018}, the authors propose a distance based distributionally robust unit commitment model (DB-DRUC) based on KL divergence, that is reformulated into a mixed integer non-linear programming problem (RDB-DRUC).
The RDB-DRUC model is scenario based and is defined as
\begin{align}
    \text{RDB-DRUC: } & \text{min } c^\intercal y + \alpha \log \left\{ \underset{s=1}{\overset{S}{\sum}}\rho_s e^{\theta_s/\alpha}\right\} + \alpha \eta \\
    s.t. \;     & \; \theta_s - q_s^\intercal x_s = 0, \forall s = 1, 2, ..., S, \\
                & \; Tx_s + Wy - h(v_s) \leq 0, \forall s = 1, 2, ..., S, \\
                & \; Ay - b \leq 0, \\
                & \; y \in \{0, 1\}^{\mathcal{M}}.
\end{align}
Here, the subscript $s$ represents the scenario index. 
$\{ v_1, \dots, v_S \}$ are uncertainty variables in the scenarios with their probabilities $\{ \rho_1, \dots, \rho_S \}$, respectively \cite{chen2018}. $\theta_s$ represents the generation cost of all units in scenario $s$, $y, x$ are the first and second stage decisions of the model \cite{chen2018}. Variable $\alpha$ is derived from the dual transformation \cite{chen2018}. 
RDB-DRUC is a large-scale mixed integer non-linear problem that can be decomposed using generalized Benders Decomposition \cite{chen2018}. The resulting subproblem can be further decomposed into a second level master- and subproblem that can be solved seperately \cite{chen2018}.
In general the RDB-DRUC model outperforms the robust UC (RUC) counterpart, as shown in the case study \cite{chen2018}. As $\eta \rightarrow \infty$, RDB-DRUC performance is equal to basic RUC, for all other values it is less conservative \cite{chen2018}. The authors further show that, by parallelizing the optimization task with four threads, the solution of the second level problem can be sped up by about $65\%$ \cite{chen2018}. From experiments with real world data it can be concluded that the model scales well even with high scenario counts \cite{chen2018}.

Parallelization as in \cite{chen2018} is not always easy to achieve. In \cite{liu2021}, a stochastic UC model for electric-gas coupled grids is proposed. Coupled systems often pose a heavy computational burden for traditional UC methods, since scenario independent variables can be correlated across scenarios when using algorithms like Benders Decomposition \cite{liu2021}. For this reason, parallelizing these algorithms is difficult. 
The authors therefore apply a modified progressive hedging (MPH) algorithm to accelerate the SUC optimization performance. The model for the electricity-gas coupled system includes wind power and gas flow scenarios. The objective function to minimize is defined as the total cost over a time horizon $t \in \mathcal{T}$ and scenarios $sc \in \mathcal{SC}$. 
Gas system constraints include gas well production limit, storage capacities, injection and withdrawal limits and storage level variation time \cite{liu2021}.
A significant optimization acceleration is achieved when compared to commercial algorithms through uniform decomposition of the stochastic model with the modified progressive hedging algorithm, which can run on parallel hardware \cite{liu2021}. In the case study, up to four workers are used in parallel \cite{liu2021}. Furthermore, instead of piecewise linear function approximation, second-order cone relaxation is applied to model gas flow \cite{liu2021}. These changes are tested in the case study.
The MPH algorithm is compared to several other optimization techniques, the most interesting for out purpose being Deterministic UC and SUC solved using Benders decomposition. While deterministic UC only needs a short amount of time to converge to a solution, the expected cost is in some cases more then ten times higher than with the other approaches \cite{liu2021}. The Benders decomposition SUC approach converges to a similar expected cost as the MPH SUC model, but it is outperformed by MPH with computational speeds being up to $17$ times faster in some cases due to the ability to be run in parallel \cite{liu2021}.

\section{Stochastic Dynamic Programming} \label{SDP_section}
In an attempt to improve UC modeling under renewable generation uncertainty through SDP, the authors in \cite{zou2019} propose a SDP based reformulation of multi-stage UC.
To optimize the model, an algorithm called Stochastic Dual Dynamic integer Programming (SDDiP) is used. SDDiP is a sampling based algorithm and is scalable to even large-scale scenario trees \cite{zou2019}. The authors propose several approaches to improve performance of SDDiP, a leveling method for Lagrangian cuts, a hybrid model approach to faster approximate the expected cost-to-go function in each stage and backwards parallelization to compute multiple scenario problems at once \cite{zou2019}.
After all the applied performance enhancements, the authors are able to show that the proposed algorithm handles multistage UC problems with a large number of scenarios \cite{zou2019}. All of the proposed improvements on the SDDiP algorithm are tested separately \cite{zou2019}. Applying all those performance improvements and running the algorithm on 32 threads ultimately results in a speedup ratio of $4.8$ and an efficiency of $15\%$ on average \cite{zou2019}.

Instead of trying to handle a large amount of scenarios through advanced algorithms, there is also the option to use probabilistic constraints instead. Nguyen et al. propose in \cite{nguyen2016} a new battery operating cost model that considers battery cycle life and charge and discharge efficiencies of batteries. To relax hard power balancing constraints, a probabilistic constraint is used that incorporates all random variables of renewable energy generation and loads \cite{nguyen2016}.
One of the main achievements of the authors is to make economic dispatch possible for several batteries without introducing more objective functions to solve to the problem \cite{nguyen2016}. The main advantage of the proposed method is that it accurately computes the actual performance of the energy storage system performance, so resources can be managed better \cite{nguyen2016}. 
The SUC is defined to minimize expected operation cost of a microgrid over a time horizon \cite{nguyen2016}. The objective function is
\begin{align}
    \min \;\; & C = \underset{k=1}{\overset{N}{\sum}}(F_k + S_k)\\
        & F_k = \mathbb{E} \{F_{g, k} + F_{b, k}^d + F_{b, c}^c + F_{m, k}\}
\end{align}
where $N$ is the time horizon, $T$ the time step, $F_k$ the total expected operation cost in period $k$, $S_k$ the total transition cost which accounts for the start-up and shut-down cost of the generators in period $k$ \cite{nguyen2016}. $F_{g, k}$ is the total operation cost of generators during period $k$, $F_{b, k}^d$ represents (dis-) charging batteries, $F_{m, k}$ is the cost due to power mismatch \cite{nguyen2016}.
The developed case study models a rural or developing nation microgrid which consists of batteries and wind, photovoltaic and diesel generators \cite{nguyen2016}. The study shows that the stochastic model can handle different load and generation scenarios well \cite{nguyen2016}. It better models the performance of energy storage systems and therefore better allocates resources that a deterministic model \cite{nguyen2016}. It is further shown that batteries provide a more cost effective choice than diesel generators, therefore increasing the energy storage system size can provide economic advantages \cite{nguyen2016}.

\section{Optimization paradigm comparison} \label{comparison_section}
In real world use cases of electricity grid balancing, RO is still the most widely used method of optimization \cite{zheng2015}. This mostly comes down to the simplicity of implementation and intuitiveness of this approach. While SP and SDP show promising results, in some cases outperforming RO \cite{hanhuawei2017}, large scenario trees and therefore long optimization times hinder wide spread adoption of these models. In addition, the complexity of the applied hedging algorithms to cope with these large scenario trees, as in \cite{zou2019} and \cite{chen2018}, can be discouraging to power generation companies.
DRO here has an advantage over the stated methods. It builds on the well researched foundations of RO, but provides more realistic results due to the use of ambiguity sets \cite{zheng2015}. The conservativeness can be controlled to some extent, as shown in \cite{yurdakul2021}, and DRO solutions do not suffer as much from inaccurate distributions of uncertain parameters as SP and SDP methods \cite{chen2018}. DRO further benefits from the fact that that the underlying probability distributions of the model parameters do not need to be determined a priori \cite{yurdakul2021}.

\section{Renewable energy outlook} \label{outlook_section}
While the reviewed papers all provide improvements to models incorporating renewable energy uncertainty, they do not take additional costs like green house gas emissions and the associated rising environmental taxes into account. This can be important, as can be seen in \cite{yurdakul2020_quantification}, where environmental unit commitment (EUC) is proposed to incorporate carbon tax payments into account. EUC therfore also minimizes green house gas emissions.
The performed case study clearly shows the advantages of EUC over CUC. It is shown that for varying carbon tax rates, EUC outperforms CUC even in scenarios where the tax is low. This is due to the increased cost of TGR generation which can only be incorporated ex-post by CUC. Applying the modifications to other models can help to more realistically predict actual generation cost of green house gas producting generation units. Yurdakul, Sivrikaya and Albayrak plan to propose models incorporating the gained experience \cite{yurdakul2020_quantification}. We conclude that research in this area is still in progress.

\section{Conclusion}
In the preceding sections, we have seen several modern stochastic approaches to the unit commitment problem. All of the discussed papers provide relevant additions to the UC problem. The main approaches we reviewed are DRO, SP and DSP. We summarized several promising methods in the field of DRO, which propose multiple improvements to (micro) grid operation, which will be increasingly important for the rising share of renewable energy. As two stage models are still the more viable and applied solution to SUC, we suggest that the DRO approach is especially worth to do further research in, since it improves the already common robust optimization through the use of ambiguity sets. It is therefore well suited to incorporate renewable generation uncertainty.
All of these advances help improving the efficiency of power production, thus enabling a more sustainable future.

\bibliographystyle{IEEEtran} 
\bibliography{papers}
\end{document}